\documentstyle[prb,aps,graphicx,multicol]{revtex}

\begin{document}
\draft
\title{Scaling behavior in a quantum wire with scatterers}
\author{Daniel Boese}
\address{Institut f\"ur Theoretische Festk\"orperphysik, 
Universit\"at Karlsruhe, D-76128 Karlsruhe, Germany \\
Forschungszentrum Karlsruhe, Institut f\"ur Nanotechnologie, 
D-76021 Karlsruhe, Germany} 
\author{Markus Lischka}
\address{Physik-Department T30, Technische Universit\"at M\"unchen, 
D-85747 Garching, Germany}
\author{L.~E.~Reichl}
\address{Center for Studies in Statistical Mechanics and Complex Systems, \\
The University of Texas at Austin, Austin, Texas 78712}
\date{\today}
\maketitle
\begin{abstract}
    We study the conductance properties of a straight two-dimensional
    quantum wire with impurities modeled by $s$-like scatterers.  Their
    presence can lead to strong inter-channel coupling.  It was shown
    that such systems depend sensitively on the number of transverse
    modes included.  Based on a poor man's scaling technique we
    include the effect of higher modes in a renormalized coupling
    constant.  We therefore show that the low-energy behavior of the
    wire is dominated by only a few modes, which hence is a way to
    reduce the necessary computing power.  The technique is
    successfully applied to the case of one and two $s$-like scatterers.
\end{abstract}

\begin{multicols}{2}

\narrowtext
\section{Introduction} \label{sec:intro}
Over the past decade, quantum transport in two-dimensional electron
systems has attracted increasing attention.  In electron waveguide
structures, the transmission and resonance phenomena have been studied
extensively, especially because the conductance was shown to be
directly related to the scattering properties of the system.  This
relation is known as Landauer-B{\"u}ttiker formula, 
\cite{landauer:57,buettiker:88,fisher:81,stone:88,baranger:89}
\begin{equation} \label{eq:landauer}
    G = \frac{e^{2}}{h} T,
\end{equation}
where $G$ denotes the 2-point conductance and $T$ the full transmission
function (spin degrees of freedom are neglected). 
Eq.~(\ref{eq:landauer}) results in a quantized conductance for
ballistic channels, which has been experimentally verified
for a large range of conduction channels.\cite{liang:99} 

It is therefore of great interest to study the properties of ballistic
wires containing impurities.  Especially for attractive scatterers the
combined effect of the scatterer itself and the backscattering off the
walls leads to interesting phenomena such as resonances and
quasi-bound
states.\cite{datta:87,bagwell:90,bagwell:92,levinson:92,kunze:92,wang:97} 
Moreover strong inter-channel coupling is introduced into the system
making
calculations quite complicated.  A simple model containing all of the
features mentioned above consists of a point-like scatterer modeled by
a two-dimensional
$\delta$-function.\cite{datta:87,bagwell:90,bagwell:92} In this paper
we present a way to dramatically reduce the computational complexity
of this problem by including the effect of the higher modes in a
renormalized coupling constant.

The outline of this paper is as follows: In the next section we
present the standard approach to calculating the conductance in this
kind of waveguide setup.  We then briefly discuss results for two
model systems for an impurity in the waveguide.  For the case of a
$\delta$-function impurity, we introduce a scaling approach and
discuss the results for the rescaled system.  We finally generalize the
scaling approach to a system with many scatterers and show results for the
case of two $\delta$-function scatterers.
\section{Conductance calculation} 
In this section we sketch the calculation of the conductance of an
electron waveguide including a general scatterer.  We restrict
ourselves to a single scatterer, because the transmission matrix for many
scatterers can be obtained by multiplying together the transmission matrices
for the individual single scatterers.

Let us first obtain a solution of the Schr\"odinger
equation for an electron in a two-dimensional waveguide with a general
scatterer.  The Hamiltonian is given by
\begin{equation} \label{eq:H}
    H = \frac{p^{2}}{2 m} + V(x,y) + V_{c}(y).
\end{equation}
$V_{c}$ represents a confinement potential restricting the transverse
movement of the electron and $V(x,y)$ represents the scattering
potential.  We assume $V(x,y)$ to be non-zero only within a finite
region small compared to the width of the channel.  We can now expand
any stationary solution $\psi_{E}(x,y)$ of the Schr\"odinger equation,
$H \psi_{E}(x,y) = E \psi_{E}(x,y)$, in a Fourier series with
$x$-dependent expansion coefficients using the complete set of
transversal modes,
\begin{equation} \label{eq:expansion}
    \psi_{E}(x,y) = \sum_{n=1}^{\infty} c_{n}(x) \chi_{n}(y),
\end{equation}
where $\chi_{n}(y)$ is the transverse state of the electron in the absence of
the scatterer. Inserting this series into the Schr\"odinger
equation and employing orthogonality of the transversal modes we
obtain a set of coupled equations,
\begin{equation} \label{eq:cmpde}
    \frac{\partial^{2}}{\partial x^{2}} c_{m} \left( x \right) +
    k_{m}^{2} c_{m} (x) = \sum_{n} M_{mn}(x) c_{n}(x) ,
\end{equation}
with the definitions of the mode coupling constants $M_{mn}(x)$ and
the wave vector $k_m$,
\begin{eqnarray} \label{eq:defV}
    M_{mn}(x) &=& \frac{2 m}{\hbar^{2}} \int \! dy \chi_m^\ast (y)
    V(x,y) \chi_n (y) ,
    \\ \label{eq:defk}
    k_m &=& \sqrt{\frac{2 m}{\hbar^{2}} \left(E - E_m\right)} .
\end{eqnarray}
Outside the scattering region the solution is given by linear combinations of
the form 
\begin{equation} \label{eq:cnsol}
    c_{n}(x) = \left\{
    \begin{array}{ll}
        A_{n} e^{i k_{n} x} + B_{n} e^{-i k_{n} x} & \quad x \ll 0 \\
        C_{n} e^{i k_{n} x} + D_{n} e^{-i k_{n} x} & \quad x \gg 0
    \end{array} \right.
\end{equation}
For $k_n$ real, we get propagating modes, for $k_n$ imaginary
evanescent modes.  By matching the wavefunctions for both regions with the 
appropriate boundary conditions, the coefficients for incoming and 
outgoing waves can be obtained.

The transmission coefficient for propagating modes is
defined as $T_{mn} = \frac{k_{n}}{k_{m}} \frac{\left| C_{n}
\right|^{2}}{\left| A_{m} \right|^{2}}$ and the total transmission
function as
\begin{equation} \label{eq:totaltrans}
    T(E) = \sum_{mn \atop \mathrm{(prop.)}} T_{mn}
\end{equation}
where the sum extends over all propagating modes. The conductance is finally
calculated using Eq.~(\ref{eq:landauer}). 

The transmission coefficients can also be obtained from the Green's
function of the system.  More specifically, it is the retarded Green's
function that governs this behavior and via its poles also describes
resonance behavior.\cite{fisher:81,taylor} The retarded Green's
function for the wire without scatterer is given by
\begin{equation} \label{eq:G0y}
    G^{0} \left( x,y,x',y' \right) = 
        \sum_{n = 1}^{\infty} \chi^\ast_n(y) \chi_n(y')
        \frac{2 m}{\hbar^{2}} \frac{e^{i k_{n}
        \left| x-x' \right|}}{2 i k_{n}}
\end{equation}
or
\begin{eqnarray}
\label{eq:G0n}
    G^{0}_{ab} \left( x,x' \right) &=&
        \int \! dy \, dy' \, \chi_a^{\ast}(y)    
        G^{0} \left( x,y,x',y' \right) \chi_{b}(y') \\
& \propto & \delta_{ab}. \nonumber
\end{eqnarray}
The full solution can then be obtained from the Dyson equation
\begin{eqnarray} 
    \lefteqn{G_{ab}(x, x')= G^0_{ab} (x,x')}  \nonumber \\ 
    &+& \sum_{c,d}\int \! dx'' \; 
    G_{ac}^0(x,x'') V_{cd} (x'') G_{db}(x'',x') \label{eq:fullG}
\end{eqnarray}
with
\begin{equation}
    V_{cd} (x'') = \int \! dy \; \chi_c^{\ast}(y) V(x'',y) \chi_{d}(y). 
\end{equation}
\section{Simple model systems}
We will now apply the general outline of the previous section to two simple
model systems: the case of the $\delta$-function scatterer described 
by 
\begin{equation} \label{eq:potdelta}
    V(x,y)=\gamma \delta(x) \delta(y-y_0)
\end{equation}
and discussed by various authors,\cite{bagwell:90,bagwell:90b,boese:00} 
as well as the slightly more realistic model with 
\begin{equation} \label{eq:potgauss}
    V(x,y)=\gamma \delta(x) \frac{1}{\sqrt{\pi} \rho} 
    \exp \left(-\frac{y^2}{\rho^2} \right)
\end{equation}
describing a softer ``impurity'' scatterer in transversal
direction.\cite{azbel:91}

For both potentials, the waveguide can be split in two separate regions 
at $x=0$ and Eq.~(\ref{eq:cnsol}) can be written as
\begin{equation} \label{eq:cnsol2}
    c_{n}(x) = \left\{
    \begin{array}{ll}
         A_{n} e^{i k_{n} x} + B_{n} e^{-i k_{n} x} & \quad x < 0 \\
         C_{n} e^{i k_{n} x} + D_{n} e^{-i k_{n} x} & \quad x > 0
    \end{array} \right.
\end{equation}
As $\psi$ must be continuous at $x = 0$ and its derivative must have a
finite jump there, the same conditions must hold for the expansion
coefficients $c_{n}(x)$.  Thus using these two conditions on
Eq.~(\ref{eq:cmpde}) with the ansatz in Eq.~(\ref{eq:cnsol2}) yields
\begin{eqnarray} \label{eq:set1}
    A_{n}+B_{n} &&= C_{n}+D_{n},
    \\
    i k_{n} \left( C_{n} - D_{n} \right) - &&
    i k_{n} \left( A_{n} - B_{n} \right) \nonumber
    \\ \label{eq:set2}
    &&= \sum_{m} M_{nm} \left( A_{m} + B_{m} \right).
\end{eqnarray}
If $\psi$ is an evanescent mode, we can set $k_{n} = i \kappa_{n}$ and
must require $A_{n} = 0$ and $D_{n} = 0$ to have a normalizable
wavefunction.  The coupling constants are given by
\begin{equation} \label{eq:Mnm}
    M_{nm}= \frac{2 m \gamma}{\hbar^2} \chi_n^{\ast}(y_0) \chi_m(y_0)
\end{equation}
for the case of the $\delta$-function scatterer, 
Eq.~(\ref{eq:potdelta}), and
\begin{equation} \label{eq:Mnm2}
    M_{nm}= \frac{2 m \gamma}{\hbar^2}
    \int \! dy \chi_m^\ast (y)
    \frac{1}{\sqrt{\pi} \rho} \exp \left(-\frac{y^2}{\rho^2} \right) \chi_n (y)
\end{equation}
for the ``impurity'' scatterer, Eq.~(\ref{eq:potgauss}). In either 
case, the coupling constants do not depend on $x$ anymore. For given 
coupling constants $M_{nm}$, the transmission coefficients can be 
computed by solving Eqs.~(\ref{eq:set1}) and (\ref{eq:set2}).
For an attractive potential ($\gamma <0$) resonance dips can be
observed just below the energy at which a new propagating mode develops.
Their position and width depends 
sensitively on the number of modes and the scatterer's
strength.\cite{boese:00}

We note that in the case of the $\delta$-function scatterer, the
integral equation for the full Green's function, Eq.~(\ref{eq:fullG}),
can be solved explicitly to yield \cite{bagwell:90b}
\begin{eqnarray} \label{eq:Gsol}
    \lefteqn{G \left( x,y,x',y' \right) = G^{0}
        \left( x,y,x',y' \right)}\qquad\qquad
    \\
    && + \frac{G^{0} \left( x,y,0,y_{0} \right) G^{0}
        \left( 0,y_{0},x',y' \right) }
        {1/\gamma - G^{0} \left( 0,y_{0},0,y_{0} \right) }.
        \nonumber
\end{eqnarray}

\begin{figure}
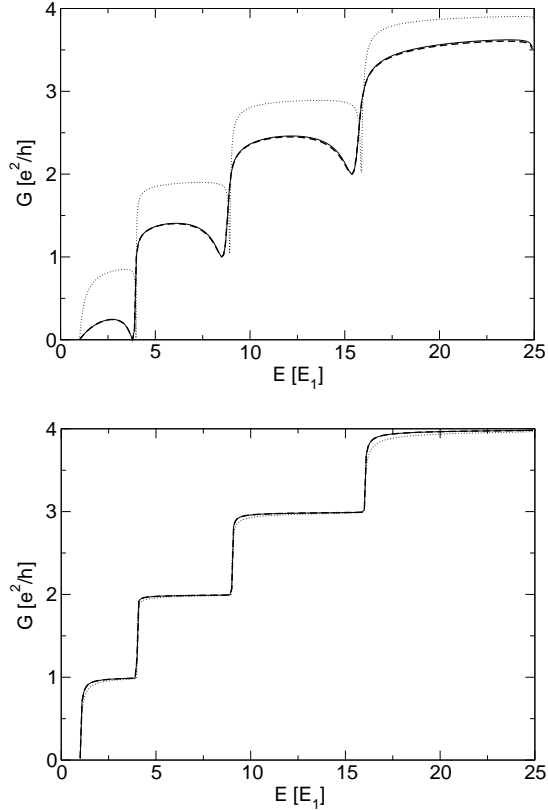

  \vbox{
    \centerline{\includegraphics[scale=0.3]{fig1a}}
    \centerline{\includegraphics[scale=0.3,angle=270]{fig1b}}
    \caption{The conductance for an attractive (top) and a
    repulsive (bottom) $\delta$-scatterer of strength  $\gamma  = \mp 7 \,
    \mathrm{feV} \, \mathrm{cm}^{2}$ and $n_c=100$ (dashed), $n_c=10$
    (dotted) and $n_c=100$ scaled down to $\tilde{n}_c=10$ (solid). The
    original curve with $n_c=100$ and the rescaled one with $\tilde{n}_c=10$
    lie nearly on top of each other.}  }
    \protect\label{figcond1}
\end{figure}

\begin{figure}
  \vbox{
    \centerline{\includegraphics[scale=0.3,angle=270]{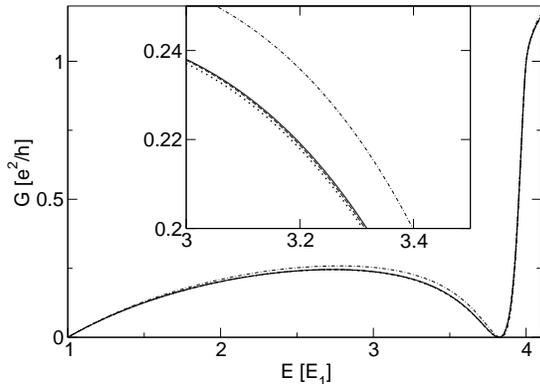}}
    \caption{The conductance for an attractive $\delta$-scatterer of
    strength $\gamma = -7 \, \mathrm{feV} \, \mathrm{cm}^{2}$,
    $n_c=100$ calculated without scaling (dotted) and with
    $\tilde{n}_{c}$ rescaled to 10 (solid), to 50 (dashed), and to 5
    (dot-dashed).  If the number of modes after the scaling is large
    enough, the difference is hardly visible.}}
    \protect\label{figcond3}
\end{figure}

\begin{figure}
    \centerline{\includegraphics[scale=0.3]{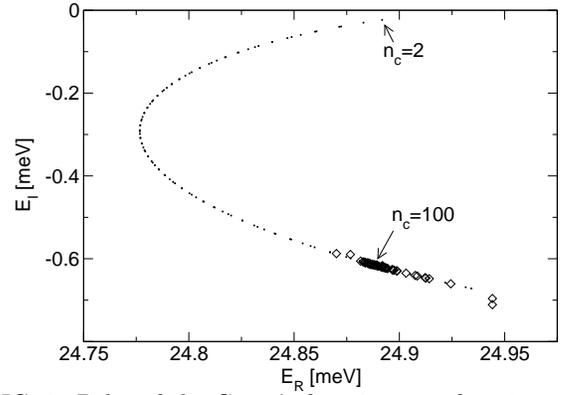}}
    \caption{Poles of the Green's function as a function of $n_c$.
    The dots indicate the pole location for values from $n_c=2$ to 
    $n_c=110$. 
    The diamonds represent the case $n_c=100$ with $\tilde{n}_c$
    varying between $2$ and $60$.}
    \protect\label{fig:poles}
\end{figure}

We first briefly summarize the results for the $\delta$-function
scatterer.  As there is no solution to a two-dimensional
$\delta$-function the problem would be ill-posed unless one restricts
the problem to a finite number of transverse modes.  The potential is
then equivalent to an $s$-like scatterer.  As more modes are included
the electron becomes more strongly localized.
For the following computations, we will use the parameters $D = 300 \,
\mathrm{\AA}$ for the width of the channel (hard wall potential),
$y_{0} = \frac{5}{12} D$ for the transversal position of the
scatterer, and the mass $m = 0.067 m_{e}$ as the effective mass of an
electron in GaAs-AlGaAs heterostructures.  The total number of modes
will be denoted by $n_c$.  For a potential strength of $\gamma = \mp 7
\, \mathrm{feV} \, \mathrm{cm}^{2}$, the resulting conductance is
shown in Figs.~\ref{figcond1} (dashed line) and \ref{figcond3} (dotted line). 
The dependence of the pole location on the number of modes can be seen in
Fig.~\ref{fig:poles} (dots), with no indication of convergence with respect
to $n_c$.

\begin{figure}
    \centerline{\includegraphics[scale=0.3]{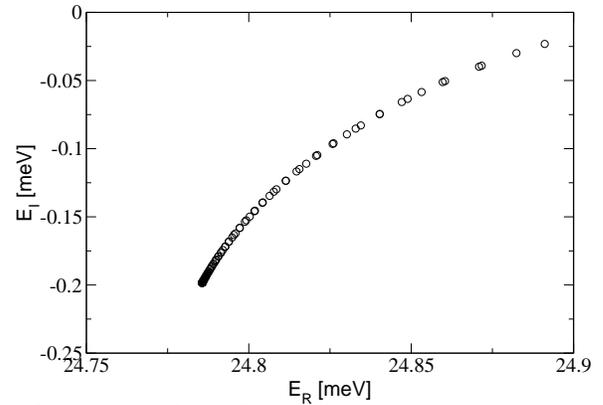}}
    \caption{Poles of the Green's function as a function of $n_c$. 
    The circles indicate the pole location for values from $n_c=2$ to
    $n_c=110$.  When using $n_c=80$ modes, the pole location is
    converged to its final position.}
    \protect\label{fig:exp}
\end{figure}

This convergence problem can be avoided by modifying the
$\delta$-function potential in $y$-direction to the smoother form of a
Gaussian function as in Eq.~(\ref{eq:potgauss}). The finite width allows the
higher modes to decouple from the lower ones. In the limit $\rho
\to 0$, we recover the $\delta$-function potential with the potential
strength $\gamma$.  By numerically evaluating the integrals for the
coupling constants, Eq.~(\ref{eq:Mnm2}), one can again obtain the full
transmission function and thus the conductance.  As expected, the
conductance curves closely resembles the results for the $\delta$-function
scatterer, showing the same characteristic resonance dip for
attractive scatterers.  Numerical results for the exact pole location
of the transmission function are shown in Fig.~\ref{fig:exp} using the
potential parameters $\gamma = - 7 \, \mathrm{feV} \, \mathrm{cm}^{2}$
and $\rho = 3.0 \, \mathrm{\AA}$.  In contrast to the
$\delta$-function potential, the result converges to its exact
solution when using a large enough, but finite number of modes.

Nevertheless, the characteristic behavior of the conductance is
identical to the $\delta$-function scatterer with a finite number of
modes.  This again demonstrates that the $\delta$-function scatterer
serves its purpose as a useful model if it is interpreted as an
$s$-like scatterer for a fixed finite number of modes.
\section{Scaling approach}
The methods presented in the last section work fine except that they require an
infinite or very large (for a strongly localized scatterer) number of modes to
be included. Simply cutting of the sum at some small number does not give
the correct answer. But calculations for large $n_c$ have the drawback that
substantial computing power is needed. Therefore it would be useful if one
could find a way to reduce the number of modes in the calculation without
neglecting their influence. A well known technique for this kind of problem is
the poor man's scaling approach.\cite{anderson} The idea is the following:
Consider a typical relevant process which includes an intermediate virtual
state. The 
sum over the virtual states shall have a cut-off. If the system exhibits
scaling behavior one can infinitesimally reduce the cut-off and incorporate
the change into a new effective coupling constant for a process not involving
these virtual states. By repeating these steps one can integrate out a large
portion of the states.

\begin{figure}
    \centerline{\includegraphics[scale=0.5]{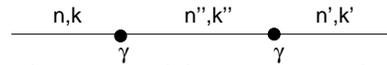}}
    \caption{A diagram describing the propagation from a state characterized
      by mode $n$ and longitudinal momentum $k$ via two scattering events into
    the state $n',k'$.}
    \protect\label{fig:diag}
\end{figure}

For the present system this means that we investigate the influence of
decreasing the mode cut-off $n_c$ on the coupling constant
$\gamma$.\cite{gamprop} In the spirit of poor man's scaling in the Kondo
problem we look at a diagram as shown in Fig.~\ref{fig:diag} which represents a
typical process in the wire. We will first discuss this in terms of a general
scatterer before we show the detailed calculation for the $\delta$-function
scatterer from above. The contribution of the diagram is given by
\begin{eqnarray}
    \sum_{c=1}^{n_c} \int \! dx'' dx''' \,G_{aa}^0(x,x'') V_{ac}(x'') 
        G_{cc}^0(x'',x''') \nonumber \\ 
    \times \, V_{cb}(x''') G_{bb}^0(x''',x') .
\end{eqnarray}
The cut-off $n_c$ stops the summation of the intermediate modes. Reducing the
cut-off by one mode amounts to a difference that equals the $n_c$'s summand.
This change should be incorporated into the new coupling constant that is taken
for a process with only one scattering event, which would have the following
form
\begin{equation}
    \int \!dx''\, G_{aa}^0(x,x'') V'_{ab}(x'') G_{bb}^0(x'',x')
\end{equation}
where the $V'_{ab}(x'')$ denotes a potential with renormalized coupling
constant. Assuming that although $n_c$ is clearly integer we can treat the
difference as infinitesimal, and that the potential can always be written as
$V_{ab}(x)=\gamma \tilde{V}_{ab}(x)$, we can formally write
\begin{eqnarray}
    \frac{\delta \gamma}{\delta n_c} &=& -\left(\tilde{V}_{ab}(x'')\right)^{-1}
    \nonumber \\ 
    &\times& \int \! dx''' V_{a n_c}(x'') G_{n_c  n_c}^0(x'',x''')
        V_{n_c b}(x''').   
\end{eqnarray}
This equation is in general difficult to solve, and could potentially contain
an explicit dependence on $a$ and $b$.\cite{abdep} We therefore show how to 
solve this equation for the $\delta$-function scatterer with a finite
number of modes as discussed above. The contribution of
the diagram is given by $\gamma G_{n_c n_c}^{0} \left(0,0\right)
\gamma$. For large $n_c$ we can thus write the change in $\gamma$ as 
\begin{equation} \label{eq:delgamma}
    \frac{\delta \gamma}{\delta n_c} = -\gamma^2 \frac{2}{D}
    \frac{2m}{\hbar^2} \sin^2 \left( \frac{\pi n_c y_0}{D} \right) 
    \frac{1}{2 i k_{n_c}} .
\end{equation}
For large $n_c$ we can approximate $k_{n_c} \approx \frac{i n_c \pi}{D}$
and integrate Eq.~(\ref{eq:delgamma}) to obtain 
\begin{eqnarray}
    -\frac{1}{\gamma} &=& \frac{2m}{\hbar^2 \pi} \left( -\frac{1}{2}
    \mathrm{Ci}\left( \frac{2 \pi y_0 n_c}{D} \right) + \frac{1}{2} \ln n_c
    \right) + \mathrm{const.}
    \nonumber \\ 
    &\cong& \frac{m}{\hbar^2 \pi} \ln n_c + \mathrm{const.}, 
\end{eqnarray}
where the last equality holds again for large $n_c$ and $\mathrm{Ci}(x)$ is
the integral cosine function. Since we expect a small change of the cut-off
not to change the result as a whole we can deduce the renormalization group
equation 
\begin{equation} \label{eq:rg}
    n_c \exp \left( \frac{\hbar^2 \pi}{m \gamma} \right) = \tilde{n}_c 
    \exp \left( \frac{\hbar^2 \pi}{m \tilde{\gamma}} \right). 
\end{equation}
With the help of Eq.~(\ref{eq:rg}) we can now calculate the
conductance for a model with a large number of modes $n_c$ by including only
$\tilde{n}_c$ modes and the renormalized coupling  
\begin{equation} \label{eq:ncscale}
    \tilde{\gamma} = \frac{1}{\frac{1}{\gamma}+\frac{m}{\hbar^2 \pi} \ln
    \frac{n_c}{\tilde{n}_c} } .
\end{equation} 
The cut-off $\tilde{n}_c$ must be chosen such that the approximations made
above are valid, i.~e.\ the renormalization procedure will eventually break
down for small $\tilde{n}_c$. Moreover, there is one strict physical criterion
for a minimum lower bound to $\tilde{n}_c$: To reproduce the resonance behavior
of the full system, it is at least necessary to include one evanescent mode in
the renormalized calculation.\cite{boese:00} We would also like to point out
that a divergence of $\tilde{\gamma}$ is not only possible in Eq.~(27), but
furthermore that it may even abruptly change sign from $-\infty$ to
$+\infty$. This is however not problematic: As can been seen from Eq.~(20) it
is $1/\gamma$ that enters the Green's function, telling us that $\gamma
\rightarrow \pm \infty$ show the same physics, independently of the sign of the
coupling strength. With increasing scatterer strength, the transmittive
behavior of the attractive scatterer indeed more and more resembles that of 
the repulsive scatterer.\cite{bagwell:90}
\section{Comparison to exact formalism} \label{sec:results}
In this section we show and compare the results when the formalism from above
is applied to the single $\delta$-function scatterer given by
Eq.~(\ref{eq:potdelta}). At the end we discuss the implications of this scaling
behavior.  In Fig.~\ref{figcond1} we show the conductance of a system with an
attractive (top) and a repulsive (bottom) $\delta$-function scatterer with a
finite number of modes. The shape of the curves is altered by a variation of
$n_c$, especially the width and position of the resonance in the attractive
case are quite sensitive (comparing e.~g. the dotted and dashed lines).  When
we apply the scaling approach, Eq.~(\ref{eq:ncscale}) to the model with
$n_c=100$ and scale it down to $\tilde{n}_c=10$, we obtain an almost identical
result (solid line). This clearly confirms the validity of our approach. In
Fig.~\ref{figcond3} we again plot the conductance for the attractive case. This
time we scale down to $\tilde{n}_c=50$, $10$, and $5$. For large $\tilde{n}_c$
the method works very well, but as we get to small values of $\tilde{n}_c$, it
eventually breaks down. However we used the assumption that $n_c$ is large in
our derivation, so this breakdown is expected. But even for $\tilde{n}_c=5$ it
still gives fairly good results. Another limit is reached when the number of
propagating modes is close to $\tilde{n}_c$. But only for $n_{\mathrm{prop}}
\geq \tilde{n}_c$ it breaks down completely, as there is no evanescent mode
left which could build up the resonance. It may seem surprising that such a
simple method yields these results. However, it confirms that the type of
process we have chosen is indeed the relevant one.

An alternative description of resonance behavior can be given in terms of
quasi-bound state. They are defined as the poles of the energy Green's
function in the complex energy plane. Hence our approach
should apply to the quantities of this approach as well.  In
Fig.~\ref{fig:poles} we show that the location of the poles can be calculated
very well with the scaling method. It is remarkable that even down to $n_c=2$,
all poles lie very close to the correct value. Moreover, they are not randomly
scattered around $n_c=100$, but rather lie on the parabolic curve of the poles
for increasing $n_c$.

Our result shows that although a true description of quantum wires with
scatterers requires the inclusion of all modes, the presence of higher modes
does not have a strong affect on the low energy behavior. Their influence can be
cast in a renormalized value of the coupling constant. Therefore one can say
that calculations done on these waveguide systems with a few number of modes
should give qualitatively correct results. This is not a trivial result,
because the mechanism of the suppression of the higher modes is not of
thermal origin. Rather it is the scaling behavior of these systems that
allows our conclusion. 
\section{Many scatterer case}
Another important consequence is that future calculations need to be done only
for a few low lying modes. This is important especially for systems with many
scatterers: Due to the scattering matrix multiplication, the computing time
grows linearly with the number of scatterers and grows more than quadratic with
the number of modes. When choosing $\tilde{n}_c$ one needs of course to
make sure, that the evanescent modes are strongly enough localized, such that
they do not carry a virtual current\cite{virtcurr} and thereby provide a minimum upper
cut-off. In order to demonstrate this, we perform a calculation for the simple
case of two $\delta$-function scatterers placed in series and separated by a
distance $d$. For a few number of modes this model has been discussed in
Reference \onlinecite{kumar}. Here we present results for the case of many
modes and discuss how they can be obtained using our scaling method.

\begin{figure}[h]
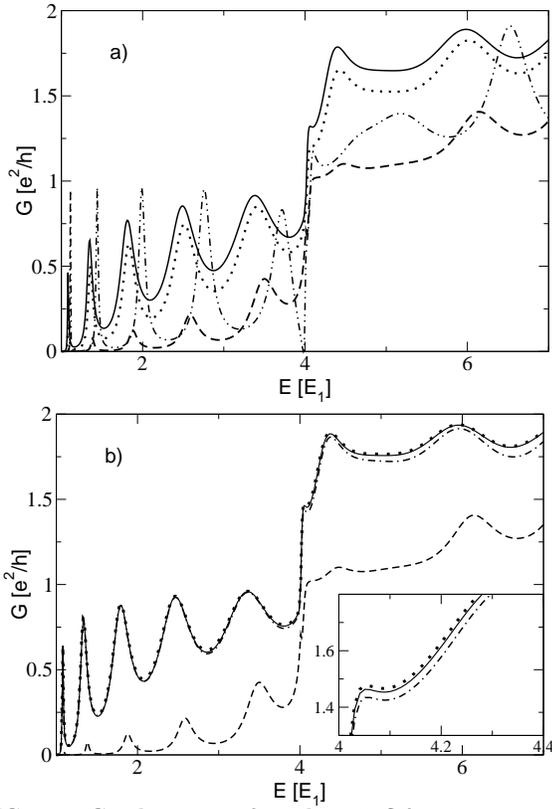

  \vbox{
   \centerline{\includegraphics[scale=0.3]{fig6a}}
    \centerline{\includegraphics[scale=0.3]{fig6b}}
    \caption{Conductance for the 2 $\delta$-function scatterer case. 
    The used parameter values are $\gamma_1 =25 \, \mathrm{feV} \,
    \mathrm{cm}^{2}$, $\gamma_2 =-25 \, \mathrm{feV} \,
    \mathrm{cm}^{2}$ and $d=90 \, \mathrm{nm}$.  In a) the different
    lines are $n_c=50$ (solid), $30$ (dotted), $10$ (dashed) and $2$
    (dot-dashed).  One can see that already from $n_{c}=10$ on, the
    qualitative behavior is the same.  \\
    In b) the scaling is applied to the $n_c=100$ case (solid) and
    performed down to $\tilde{n}_{c}=10$ (dotted)
    and $2$ modes (dot-dashed) .  The dashed line shows how
    well it compares to the case with $n_{c}=10$ modes without
    scaling.} }
    \protect\label{fig:2delta}
\end{figure}

The specific system under consideration consists of two $\delta$-function
scatterers located at $(0,y_1)$ and $(d,y_2)$, with potential strengths
$\gamma_1$ and $\gamma_2$ respectively. The waveguide itself shall be the same
as before. The Green's function of this system can be obtained analytically and
is given by the lengthy, however very useful expression
\begin{eqnarray}
    G(x,y,x',y') &=& G^0(x,y,x',y') + \gamma_1 G^0(x,y,0,y_1)
        \frac{c_1+a_1c_2}{1-a_1 a_2} 
    \nonumber \\
    &+& \gamma_2 G^0(x,y,d,y_2)
        \frac{c_2+a_2c_1}{1-a_1 a_2} 
\end{eqnarray}
where the following abbreviations have been used:
\begin{eqnarray}
    c_1 &=& \frac{ G^0(0,y_1,x',y')}{1-\gamma_1  G^0(0,y_1,0,y_1)}, \nonumber\\
    c_2 &=& \frac{ G^0(d,y_2,x',y')}{1-\gamma_2  G^0(d,y_2,d,y_2)}, \nonumber\\
    a_1 &=& \frac{ \gamma_2 G^0(0,y_1,d,y_2)}{1-\gamma_1  G^0(0,y_1,0,y_1)},
    \nonumber \\ 
    a_2 &=& \frac{ \gamma_1 G^0(d,y_2,0,y_1)}{1-\gamma_1  G^0(d,y_2,d,y_2)}.
\end{eqnarray}
To apply our scaling method, we employ the approximation that the evanescent
modes fall off exponentially. Hence contributions to the scaling of order
$\gamma_1 \gamma_2$ are strongly suppressed for higher modes, i.~e. $|k_n d|$
is large.  We can then simply use the single scatterer result and apply it
independently to both coupling constants. The results are shown in
Fig.~\ref{fig:2delta}~a) and b).

In Fig.~\ref{fig:2delta}~a) we compare the conductance for various number of
modes. Already for two modes the peaks in the lowest subband with one
evanescent mode are present. The inclusion of higher modes however leads to a
strong shift in position and shape. Although the true peak structure becomes
visible for $n_c=10$, it is fully developed only for many more modes. In
addition, the results in the second 
subband are completely wrong for two modes. This should be expected since no
evanescent mode is present. In Fig.~\ref{fig:2delta}~b) we compare our scaling
result to the exact one and again find that it works very well.  It is
remarkable that already for two modes the deviation in the first subband is
negligible.

The generalization to many scatterers is obvious: one only has to keep in mind
to take enough evanescent modes to mimic the decaying currents between the
scatterers. The Fabry-Perot behavior that causes the oscillations is created by
the propagating modes, however it can be enhanced by the first few evanescent
ones.
\section{Conclusion} \label{sec:conc}
In this article we discussed the conductance properties of ballistic quantum
wires containing impurities.  As shown, they depend strongly on the number of
modes included and the calculations can thus become quite difficult.  We proved
that it is possible to incorporate the effect of higher modes into an effective
coupling constant for the lower lying modes.  This was done by a poor man's
scaling approach.  For the $\delta$-function scatterer with a fixed number of
modes we compared our result to exact calculations and found very good
agreement.  We therefore conclude that our approach does not only allow for
quicker calculations but also shows that the system's behavior is governed by a
few low-lying modes.
\section*{Acknowledgments}
We thank Herbert Schoeller and Michele Governale for useful
discussions. D.\ B.\ acknowledges financial support from the 
DFG Graduiertenkolleg 284:``Kollektive Ph\"anomene im
Festk\"orper''. L.\ E.\ R.\ acknowledges partial support from Welch Foundation
Grant No.\ 1052 and U.S.\ DOE Grant No.\ DE-FG03-94ER14405.

%
\vspace{10cm}
%
%

\end{multicols}


\begin{references}
\bibitem{gamprop}
Regardless of the scattering potential, $V(x,y)$ can always be cast in a form
such that it is proportional to a coupling constant $\gamma$.
\bibitem{abdep}
The explicit $a$ and $b$ dependence can in principle be treated by carrying
them along all the way. This however would lead to a serious complication of
the scaling and the conductance calculations.
\bibitem{virtcurr}
We refer to a virtual current as an evanescent mode that decays from one
scatterer but 'reaches' the other scatterer and therefore can carry a current.
\end{references}
\end{document}